\documentclass[useAMS,usenatbib]{mn2e}
\usepackage{graphics,epsfig,lscape,txfonts,color}
\usepackage[english]{babel}
\usepackage{longtable}

\newcommand{\mr}{\mathrm}

\newcommand{\hcm}[1]{$\times 10^{#1}$ cm$^{-2}$}

\newcommand{\ergcm}[1]{$\times10^{#1}$~\hbox{erg~cm$^{-2}$~s$^{-1}$}}

\newcommand{\ergs}[1]{$\times10^{#1}$~\hbox{erg~s$^{-1}$}}

\newcommand{\lb}{\left}
\newcommand{\rb}{\right}
\def\eg{e.\,g.}                                      

\def\maxi{MAXI~J1543-564}

\title[MAXI~J1543-564: 2011 outburst]{The faint 2011 outburst of the black hole X-ray binary candidate MAXI~J1543-564}
\author[H. Stiele, T. Mu\~noz-Darias, S. Motta, and T. M. Belloni]{H. Stiele$^{1}$\thanks{E-mail:
holger.stiele@brera.inaf.it}, T. Mu\~noz-Darias$^{2}$, S. Motta$^{1,3}$, and T. M. Belloni$^{1}$\\
$^{1}$INAF-Osservatorio Astronomico di Brera, Via E. Bianchi 46, I-23807 Merate (LC), Italy \\
$^{2}$School of Physics and Astronomy, University of Southampton, Southampton, Hampshire, SO17 1BJ, United Kingdom\\
$^{3}$Universit\`a dell`Insubria, Via Valleggio 11, I-22100 Como, Italy}
\begin{document}

\date{2012 January 27}

\pagerange{\pageref{firstpage}--\pageref{lastpage}} \pubyear{2011}

\maketitle

\label{firstpage}

\begin{abstract}
We report on a spectral-timing analysis of the black hole X-ray binary candidate MAXI J1543-564 during its 2011 outburst. All 99 pointed observations of this outburst obtained with the Rossi X-ray Timing Explorer (RXTE) were included in our study. We computed the fundamental diagrams commonly used to study black hole transients, and fitted power density and energy spectra to study the spectral and timing parameters along the outburst. The determination of timing parameters and hence of exact transitions between different states was hampered by the rather low count rate at which his outburst was observed.  We detected two periods of exponential decay, one after the source was brightest, which was interrupted by several flares, and another one during the high/soft state. The detection of these decays allowed us to obtain an estimate for the source distance of at least 8.5 kpc. This leaves two possible explanations for the observed low count rate; either the source has a distance similar to that of other black hole X-ray binary candidates and it is intrinsically faint, or it has a similar luminosity, but is located more than 12 kpc away from us. Furthermore, in the high/soft state the  source spectrum appears to be completely disc dominated. 
\end{abstract}


\section{Introduction}
Apart from a few persistent systems, most black hole X-ray binaries (BHT) are transient. They spend most of their time in quiescence, in which they are too faint to be detectable with present X-ray instruments \citep[see e.\,g.][]{1998ASPC..137..506G}. But from time to time they go into outburst. The peak luminosities reached during outburst are broadly distributed around about 0.1 -- 0.2 of the Eddington luminosity \citep{1997ApJ...491..312C,2010MNRAS.403...61D}. Taking the luminosity ratios and black hole masses for the sample of X-ray binaries studied by \citet{2006MNRAS.370..837G}, one obtains peak luminosities (1.5 -- 12 keV) in the range of 1.5\ergs{37} to 8\ergs{38}. It should be noted that the same source can show outbursts at rather different peak luminosities, examples are XTE J1550-564 \citep{2006MNRAS.370..837G} or GX 339-4 \citep{2011MNRAS.418.2292M}. Nowadays there is general agreement that these outbursts begin and end in the low hard state (LHS) and that there is in between a transition to the high soft state (HSS), where the back transition happens at about 0.1 of the luminosity reached during the outward transition \citep{2010MNRAS.403...61D}. The different states through which a BHT evolves during an outburst can be identified with the help of the hardness intensity diagram \citep[HID;][]{2001ApJS..132..377H,2005A&A...440..207B,2005Ap&SS.300..107H,2006MNRAS.370..837G,2006csxs.book..157M,2009MNRAS.396.1370F,2010LNP...794...53B,2011BASI...39..409B}, the hardness-rms diagram \citep[HRD;][]{2005A&A...440..207B}, and the rms-intensity diagram \citep[RID;][]{2011MNRAS.410..679M}. The spectrum of the HSS is clearly dominated by an optically thick, geometrically thin disc \citep{1973A&A....24..337S}. This state shows very little rapid variability \citep[fractional rms $\sim$1 per cent, e.g.][]{2005A&A...440..207B}. In contrast, rms of several tens of per cent is observed in the LHS and the emission is dominated by thermal Comptonization in a hot, geometrically thick, optically thin plasma located in the vicinity of the black hole, where softer seed photons coming from an accretion disc are up-Comptonized \citep[see][for recent reviews]{2007A&ARv..15....1D,2010LNP...794...17G}. However, the nature of the Comptonizing medium is still a hot topic of ongoing discussion 
\citep[\eg][]{1998MNRAS.301..435Z,2005ApJ...635.1203M,2010ApJ...717.1022D}. Furthermore, the exact properties and mechanism of the transitions between different states are still under debate. We follow here the classification given in \citeauthor{2010LNP...794...53B} (2010; see also \citealt{2005A&A...440..207B,2005Ap&SS.300..107H}), which comprises -- apart from the LHS and HSS -- a soft as well as a hard intermediate state (SIMS/HIMS); see however \citet{2006csxs.book..157M} for an alternative classification and \citet{2009MNRAS.400.1603M} for a comparison. 

In addition to this long-term variability (on timescales of several month up to years), very rapid (on sub-second timescales), non-periodic variability is observed. A common feature in almost all BHTs are low-frequency quasi-periodic oscillations (LFQPOs) with frequencies ranging from a few mHz to $\sim$10 Hz. Three different types of LFQPOs can be distinguished \citep{2005ApJ...629..403C,1999ApJ...526L..33W}, and their presence can be related to different states \citep{2011BASI...39..409B}. In the LHS and HIMS type-C QPOs are observed, while type-B QPOs are only observable during SIMSs. Type-A QPOs can be observed sometimes in the HSS.

\maxi\ was discovered by MAXI/GSC \citep[the Monitor of All-sky X-ray Image / Gas Slit Camera;][]{2009PASJ...61..999M}, which is installed on the International Space Station (ISS), on 2011 May 08 \citep{2011ATel.3330....1N}. The outburst of this source has been followed up by MAXI, Swift \citep[\eg][]{2011ATel.3336....1K,2011ATel.3331....1K}, and RXTE \citep[Rossi X-ray Timing Explorer; \eg][]{2011ATel.3334....1A,2011ATel.3355....1M}. The detection of type-C QPOs, together with a decrease in fractional rms as well as hardness ratio and the steepening of the photon index led to the classification of \maxi\ as a black hole candidate (BHC) X-ray binary \citep{2011ATel.3341....1M}. The detection of a radio counterpart with an apparently optically thin spectrum was reported in \citet{2011ATel.3364....1M}, while the detection of an optical counterpart remains questionable, as the possible counterparts do not show detectable variability of their optical emission \citep{2011ATel.3359....1R,2011ATel.3365....1R,2011ATel.3372....1R}.

Here, we study in detail the evolution of the spectral and timing properties of the source along its 2011 outburst. Although the analysis is hampered by the rather low count rate at which this outburst is observed, the overall bahaviour of the spectral and timing properties suggest this source to be a black hole X-ray binary.

\section[]{Observations}
We analysed 99 RXTE observations of \maxi\ performed between 2011 May 10 and September 30. The variability study presented in this paper is based on data from the Proportional Counter Array (PCA). We computed power density spectra (PDS) for each observation following the procedure outlined in \citet{2006MNRAS.367.1113B}. We limited PDS production to the Proportional Counter Array (PCA) channel band 0 -- 35 (2 -- 15 keV) and used 16 second long stretches of \textsc{GoodXenon}, \textsc{Event} and  \textsc{SingleBit} mode data.

The PCA Standard 2 mode (STD2), which covers the 2--60 keV energy range with 129 channels, was used for the spectral analysis. We determined hardness ratio using channels 7 -- 13 (2.87 -- 5.71 keV) for the soft band, and channels 14 -- 23 (5.71 -- 9.51 keV) for the hard band. For each observation background and dead-time corrected energy spectra were extracted using the standard RXTE software within \textsc{heasoft} V.~6.9. For the spectral fitting, Proportional Counter Unit (PCU) 2 was solely used. To account for residual uncertainties in the instrument calibration a systematic error of 0.6 per cent was added to the PCA spectra\footnote{A detailed discussion on PCA calibration issues can be found at: http://www.universe.nasa.gov/xrays/programs/rxte/pca/doc/rmf/pcarmf-11.7/}. We had to exclude all High Energy X-ray Timing Experiment (HEXTE) data from our analysis, since most of the HEXTE spectra contain strong residuals that are related to the difficulties in determining the background contribution in the spectra since the ``rocking" mechanism of HEXTE is broken \citep[see also][]{2011MNRAS.tmp.1664S}.

\section[]{Analysis and Results}
\subsection[]{Fundamental diagrams}
\begin{figure}
\resizebox{\hsize}{!}{\includegraphics[clip,angle=-90]{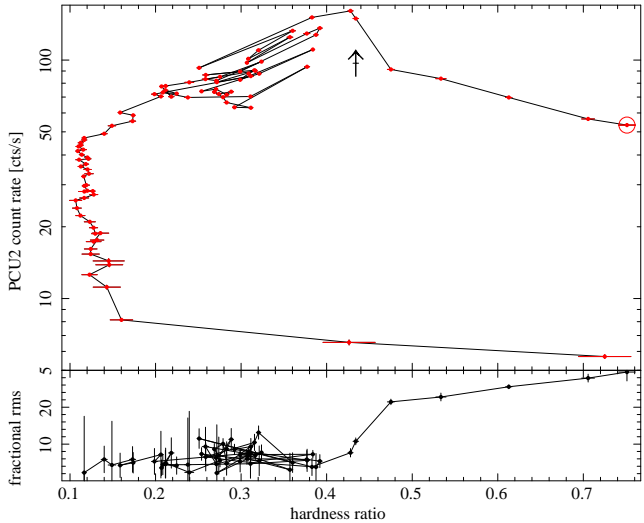}}
\caption{Upper panel: Hardness intensity diagram of the whole outburst, obtained using RXTE observations. Intensity corresponds to the count rate within the STD2 channels 0 -- 31 (2 -- 15 keV) and hardness is defined as the ratio of counts in 7 -- 13 (2.87 -- 5.71 keV) and 14 -- 23 (5.71 -- 9.51 keV) STD2 channels. Each point represents an entire observation. Consecutive observations are joined by a solid line. An arrow marks the X-ray observation located closest to the radio observation reported in \citet{2011ATel.3364....1M}. Lower panel: corresponding hardness-rms diagram within the 0.1 -- 64 Hz frequency band. Due to the low count rate and variability levels, the rms is unconstrained after observation \#59 and hence it is not shown in the diagram.}
\label{Fig:HID}
\end{figure}

The HID, shown in Fig.\,\ref{Fig:HID}, reveals that the outburst of \maxi\ took place at a rather low count rate compared with outbursts of other black hole X-ray binary candidates, such as \eg\ H1743-322 \citep{2009ApJ...698.1398M}, XTE J1650-500 \citep{2004NuPhS.132..416R}, GX 339-4 \citep{2011MNRAS.tmp.1664S}, or XTE J1752-223 \citep{2010ApJ...723.1817S,2010tsra.confE..32S}. In addition, we computed the HRD (Fig.\,\ref{Fig:HID}) and the RID (not shown).\@ The fractional rms was computed within the frequency bands 0.1--64 Hz, and 0.1--16 Hz, following \citet{1990A&A...227L..33B}. Due to the low count rate and variability levels, the rms is unconstrained after observation \#59 and hence it is not considered in the analysis (and in the HRD).

\maxi\ describes the standard q-shaped pattern in the HID moving from observation \#1 (encircled dot in Fig.\,\ref{Fig:HID}) in counter clockwise direction. It is not easy to decide whether the first RXTE observation still belongs to the LHS, or already corresponds to the HIMS, as the initial flux rise was not observed by RXTE. The HRD and the RID are not particularly useful in this study given our poor constraints on the rms during the major part of the outburst. The RID does not give any further insight as observation \#1 can correspond to the last point on the hard line \citep[i.e. sharp, linear rms-flux relation;][]{2011MNRAS.410..679M}  as well as to the first point of the turn-off from the hard line. During the first five observations a monotonic decrease in fractional rms from 29.2$\pm$2.3\% to 20.5$\pm$0.7\% takes place. In the corresponding power density spectra strong type-C QPOs are visible. In observation \#7, in which the source is brightest, the fractional rms is 7.6$\pm$1.1\%, so within the 5--10\% range, in which type-B QPOs are expected \citep{2011MNRAS.410..679M,2011MNRAS.415..292M}. However, no QPO is detected in this observation. We obtain a 3$\sigma$ upper limit of 4.11\% rms for the amplitude.

\begin{figure}
\resizebox{\hsize}{!}{\includegraphics[clip]{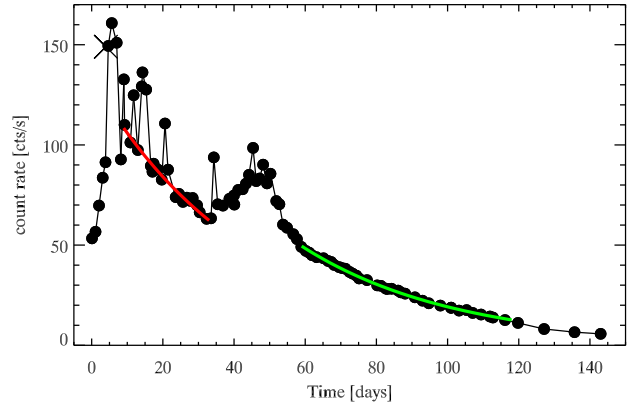}}
\caption{Light curve (2 -- 15 keV) of the recent outburst of \maxi. The time of the first observation (55691.1 MJD) was selected as T$=0$. An X marks the time of the radio observation reported in \citet{2011ATel.3364....1M}. Over plotted are two broader lines indicating the two exponential decays. They are derived from fitting the function $C(t)=A*\exp\lb(-\lb(t-t_0\rb)/\tau\rb)$ to each of the datasets. For the decay during the HSS (marked in green/light grey) we obtained the following parameters: decay time $\tau\simeq43$ days and $A\simeq18$ cts/s, $t_0\simeq101$ days. For the decay during the flaring period (marked in red/dark grey) we assumed the same decay time as observed in the HSS and obtained $A\simeq15$ cts/s, $t_0\simeq94$ days.}
\label{Fig:LC}
\end{figure}

The source then enters into a phase of the outburst where the hardness ratio stays mainly between 0.35 and 0.19. Interestingly, there are some excursions with hardness ratios of 0.35 to 0.40, which are associated to flares (see Fig.\,\ref{Fig:LC}). In systems like MAXI J1659-152 and specially GX 339-4 \citep{2011MNRAS.415..292M,2011MNRAS.418.2292M} we showed that similar peaks tended to be associated with the SIMS, as they showed type-B QPOs \citep[see also][]{2004MNRAS.355.1105F}. But again, no type-B QPOs are detectable for \maxi. Thus, we cannot clearly identify a SIMS. 
The fade through the HSS takes place at rather constant hardness ratio, so it does not show strong color variations as observed in other systems (such as \eg\ H 1743-322 or XTE J1752-223). The penultimate observation can be associated with the HIMS on the decay branch, while the last observation seems to indicate that the source has returned to the LHS.

Apart from the flares, the light curve reveals two exponential decays, which are marked with coloured lines in Fig.\,\ref{Fig:LC}. The first decay takes place during the intensity decrease after the source was brightest and is interrupted by several flares. The second one happens during the HSS when the count rate decreases at more or less constant hardness ratio. We fitted the decay during the HSS with an exponential. The resulted best-fit is indicated by a green (light grey) line in Fig.\,\ref{Fig:LC}, and we obtained a decay timescale of $\sim$43 days.  
Since the X-ray decay light curve is subject to additional variations beyond that described by the model, the fit only globally describes the decay with a rather large $\chi^2$ (255 for 34 degrees of freedom). We also fitted the decay during the flaring period with an exponential assuming that the decay timescale is the same as in the HSS. 
The red (dark grey) line in Fig.\,\ref{Fig:LC} visualises the best-fit to the exponential decay during the flaring period.

\subsection[]{Spectral evolution}
\begin{figure}
\resizebox{\hsize}{!}{\includegraphics[clip]{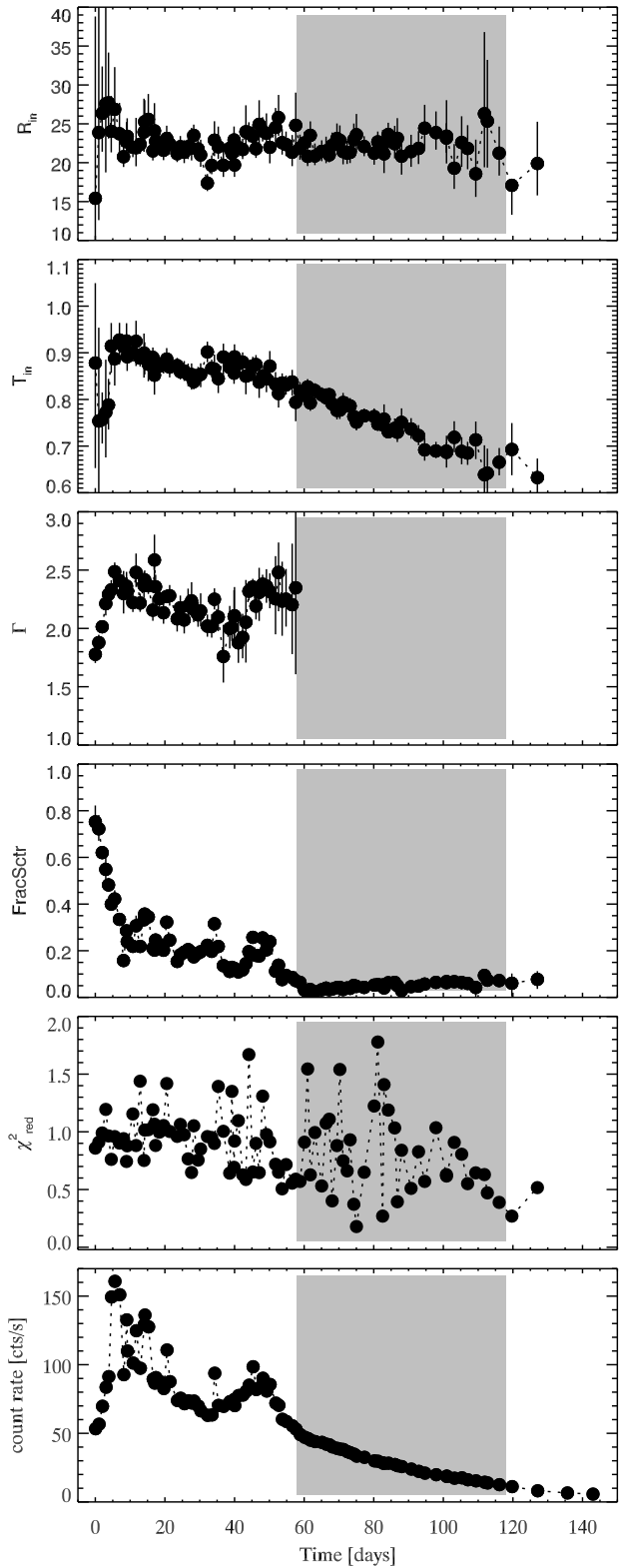}}
\caption{Temporal evolution of the different spectral parameters. The parameters shown are inner disc radius, inner disc temperature, photon index, fraction of up-scattered radiation and reduced $\chi^2$. The time of the first observation (55691.1 MJD) was selected as T$_0$. The exponential decay during the HSS is indicated by the grey shaded area. For observations for which the photon index is fixed to the value observed from the GRXE ($\Gamma\sim$2.1) no photon index is indicated.}
\label{Fig:SpecPar}
\end{figure}

We extracted energy spectra for each observation and fitted them with \textsc{isis} V.~1.6.1 \citep{2000ASPC..216..591H}. We uniformly fitted PCA spectra in the 3 --  40 keV range with a partially Comptonized multi-color disc blackbody model, including foreground absorption. The disc emission was approximated by the \emph{diskbb} model \citep{1984PASJ...36..741M} and the \emph{simpl} model \citep{2009PASP..121.1279S} was used for Compton scattering. The latter one being an empirical convolution model that converts a given fraction of the incident spectrum into a power law shape with a photon index $\Gamma$. If needed a Gaussian was added to account for the excess at 6.4 keV. For the foreground absorption we used the \emph{TBabs} model \citep{2000ApJ...542..914W}, with $N_{\mr{H}}$ fixed to 1.4\hcm{22} \citep{2011ATel.3662....1K}. We also tried $N_{\mr{H}}$ fixed to 0.9\hcm{22} \citep{2011ATel.3331....1K} . The results of the fits for $N_{\mr{H}}=$1.4\hcm{22} are shown in Fig.\,\ref{Fig:SpecPar}. The increased foreground absorption leads to a slightly enlarged inner disc radius and to a slightly reduced inner disc temperature. However, all values are consistent within errors for both parameters. The radius of the inner disc stays rather constant throughout the whole outburst. The photon index increases during the first seven observations, and throughout the remaining part of the outburst the temperature at the inner edge of the accretion disc decreases. Values can be found in Table \ref{Tab:Val}.  
We would like to point out that the values of the disc parameters should be taken with care, as the working range of PCA (3 -- 40 keV) allows to see only the high energy part of the disc black body component, above the Wien peak. The missing coverage of lower energies might also lead to the detection of a constant inner disc radius. In addition, it is known that the spectral parameters derived from the \emph{diskbb} model should not be interpreted literally \citep[see eg][]{2000MNRAS.313..193M,2006ARA&A..44...49R}. 

\begin{table}
\caption{Spectral parameters for two different values of foreground absorption. For the photon index the limiting values of the initial increase (obtained from observation \#1 and \#7) are given. The inner disc temperature decreases from observation \#8 throughout the outburst (obs.\ \#97). Values for each observation obtained with $N_{\mr{H}}=$1.4\hcm{22} are given in Table \ref{tab:data_s}.}
\begin{center}
\begin{tabular}{ccc}
\hline\noalign{\smallskip}
Parameter &$N_{\mr{H}}=$0.9\hcm{22}  & $N_{\mr{H}}=$1.4\hcm{22}  \\
\hline\noalign{\smallskip} 
$R_{\mr{in}}$ [km] & $\sim$ 20 & $\sim$ 22 \\
$T_{\mr{in}}$ [keV]& $0.96^{+0.04}_{-0.05}\to0.64\pm0.04$ & $0.93^{+0.04}_{-0.05}\to0.63\pm0.04$ \\
$\Gamma$ & $1.77\pm0.07\to2.46^{+0.08}_{-0.07}$& $1.78\pm0.07\to2.49^{+0.08}_{-0.07}$ \\
\hline\noalign{\smallskip}
\end{tabular} 
\end{center}
\label{Tab:Val}
\end{table}

\begin{table*}
\caption{Spectral parameters derived from the best fit for each observation. A model consisting of simpl and diskbb was used. The foreground absorption was fixed at $N_{\mr{H}}=$1.4\hcm{22}.This is a sample of the full table, which is available with the electronic version of the article. See Supporting Information.}
\begin{center}
\begin{tabular}{lrrrrlccccc}
\hline\noalign{\smallskip}
 \multicolumn{1}{c}{\#} &\multicolumn{1}{c}{Obs. id.} &  \multicolumn{1}{c}{day} &  \multicolumn{1}{c}{MJD} & \multicolumn{1}{c}{$\chi^2_{\mr{red}}$} & \multicolumn{1}{c}{diskbb norm} & \multicolumn{1}{c}{Tin} & \multicolumn{1}{c}{$\Gamma$} &  \multicolumn{1}{c}{FracSctr}\\
  & & & & & & \multicolumn{1}{c}{[keV]} & &\\
\hline\noalign{\smallskip} 
 1   & 96371-02-01-00    &     0.0 & 55691.1 & 0.86  & $  118.92_{- 68.92}^{ +633.85}    $&$  0.88_{-0.23 }^{+0.17}   $&$  1.78_{-0.07 }^{+0.07}   $&$ 0.752_{-0.051}^{+ 0.070} $\\
 2   & 96371-02-01-01    &     1.0 & 55692.1 &  0.91  & $  285.02_{-205.50}^{+958.76}    $&$  0.75_{-0.16 }^{+0.20}   $&$  1.88_{-0.07 }^{+0.07}   $&$ 0.723_{-0.051 }^{+0.059} $\\
 3   & 96371-02-01-02    &     2.0 & 55693.1 &  0.99  & $  347.48_{-118.20}^{+176.81}    $&$  0.76_{-0.05 }^{+0.06}   $&$  2.01_{-0.04 }^{+0.04}   $&$ 0.620_{-0.021 }^{+0.022} $\\ 
 4   & 96371-02-02-00    &     3.0 & 55694.1 &   1.19  & $  380.05_{-204.17}^{+456.35}    $&$  0.77_{-0.10 }^{+0.11}   $&$  2.21_{-0.10 }^{+0.09}   $&$ 0.549_{-0.049 }^{+0.054} $\\  
 5   & 96371-02-02-01    &     3.8 & 55694.9 &   0.96  & $  383.38_{-116.40}^{+200.46}    $&$  0.79_{-0.05 }^{+0.06}   $&$  2.29_{-0.07 }^{+0.08}   $&$ 0.482_{-0.027 }^{+0.031} $\\ 
 6   & 96371-02-02-02    &     4.6 & 55695.7 &   0.76  & $  288.73_{- 61.31}^{+  120.16}   $&$  0.91_{-0.06 }^{+0.05}   $&$  2.33_{-0.08 }^{+0.08}   $&$ 0.400_{-0.024}^{+ 0.031} $\\ 
 7   & 96371-02-02-03    &     5.6 & 55696.7 &  0.96  & $  361.01_{- 74.02}^{+ 160.49}    $&$  0.89_{-0.06 }^{+0.04}   $&$  2.49_{-0.07 }^{+0.08}   $&$ 0.422_{-0.024}^{+ 0.037} $\\  
\hline\noalign{\smallskip}
\end{tabular} 
\end{center}
\label{tab:data_s}
\end{table*}

For observations taken after day 58 (2011 July 7), the significance to detect photons at higher energies decreases and above 20 keV the spectrum becomes flat. According to the results presented in \citet{2010A&A...512A..49T} the Galactic ridge X-ray emission (GRXE) can be approximated by a cut-off power law with $\Gamma \simeq$0.0 in the 20 -- 80 keV range. This means that the channels corresponding to higher energies detect only the GRXE, as PCA is a non-focusing instrument. The observable source spectrum should be purely disc dominated, as the faint emission of the hot, Comptonized plasma will be barely distinguishable from the GRXE. This is supported by the fact, that we cannot obtain decent constrains for the slope of the power law. Therefore, we limited our spectral investigations to the 3 -- 10 keV range and fixed the photon index at the value observed from the GRXE \citep[$\Gamma\sim2.1$;][]{2006A&A...452..169R} for all observations taken after this date. The fixed photon index is not indicated in Fig.\,\ref{Fig:SpecPar}.

Despite the detectability of the source in the last two observations, we exclude them from our spectral investigations due to their low count rates. It is very difficult to securely disentangle the source spectrum from the background spectrum and it is impossible to clearly constrain the presence and contribution of different spectral components, such as accretion disc and hot plasma, to the overall emission.  

For the first two observations it is possible to obtain acceptable fits with an absorbed power law plus Gaussian model, meaning that no additional disc component is needed. The photon indices found are -- depending on the assumed foreground absorption ($N_{\mr{H}}=$0.9/1.4\hcm{22}) -- 1.82$^{+0.02}_{-0.08}$/1.85$^{+0.02}_{-0.05}$ and 1.96$^{+0.02}_{-0.03}$/1.99$^{+0.02}_{-0.03}$, respectively.
 
\begin{figure}
\resizebox{\hsize}{!}{\includegraphics[clip]{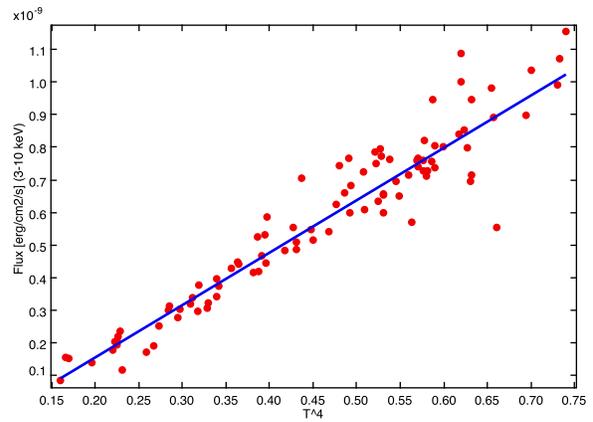}}
\caption{Correlation between soft disc flux in the 3 -- 10 keV range and the fourth power of the temperature.}
\label{Fig:FluxTemp}
\end{figure}

Figure \ref{Fig:FluxTemp} shows the linear correlation between the flux of the soft disc in the 3 -- 10 keV range and the fourth power of the corresponding temperature at the inner edge of the accretion disc. This behaviour has to be expected, as we observe a constant inner disc radius. The first two observations for which we obtained acceptable fits without including a disc component are not shown in this plot. 

\subsection[]{Quasi periodic oscillations}
\begin{figure}
\resizebox{\hsize}{!}{\includegraphics[clip, angle=-90]{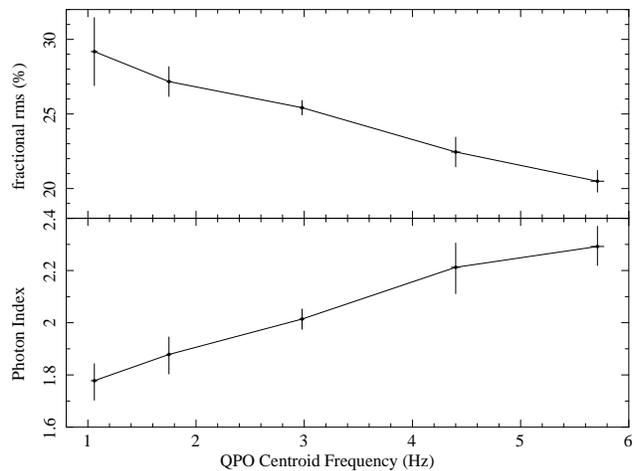}}
\caption{Upper panel: total fractional rms versus QPO centroid frequency relation of the first five observations, which showed strong type-C QPOs. Lower panel: Photon index versus QPO centroid frequency relation for the same observations.}
\label{Fig:QPO}
\end{figure}

We have also studied the evolution of the main QPO properties. The only QPOs that have been detected are of type-C and they have been found in the first five RXTE observations. After subtracting the contribution due to Poissonian noise \citep{1995ApJ...449..930Z} the PDS were normalised according to \citet{1983ApJ...272..256L} and converted to square fractional rms \citep{1990A&A...227L..33B}. PDS fitting was carried out with the standard \textsc{xspec} fitting package \citep{1996ASPC..101...17A} by using a one-to-one energy-- frequency conversion and a unit response. The noise components as well as the QPO feature have been fitted with Lorentzians, following \citet{2002ApJ...572..392B}. The behaviour of type-C QPOs is similar to that observed in other BHTs \citep[see \eg][]{2010LNP...794...53B}. We see the centroid frequency increasing with hardness. Following \citet{2004A&A...426..587C} and \citet{2011MNRAS.418.2292M} we have plotted total rms as a function of the QPO frequency (Fig.\,\ref{Fig:QPO}). As it was found in those works, the type-C QPOs follow a clear negative correlation. In addition, we have plotted the photon index obtained form our spectral investigation as a function of the QPO frequency (Fig.\,\ref{Fig:QPO}). The obtained positive correlation was expected \citep[see \eg][]{2003A&A...397..729V,2007ApJ...663..445S,2009ApJ...699..453S}.

\section[]{Discussion}
The evolution of \maxi\ during its 2011 outburst is consistent with that usually observed from black hole X-ray binaries. The hardness-intensity diagram is rather typical, with a hard-to-soft transition, a flux decay during a soft (accretion-disc-dominated) state and a final soft-to-hard transition towards quiescence. Unusual is the low count rate at which the outburst takes place. Comparison with other black hole candidates, such as MAXI J1659-152, GX 339-4, H 1743-322, or XTE J1752-223, reveal that the hard-to-soft transition in \maxi\ takes place at count rates normally observed during the soft-to-hard transition. During the soft state the count rate decreases dramatically, reaching values comparable to the background count rate. The low count rates lead to enhanced uncertainties in the determination of the timing properties. Thus we cannot establish clear boundaries between different states. We observe a decrease in fractional rms and an increase in the centroid frequency of type-C QPOs during HIMS. However, it remains an unsolved question whether the first observation belongs still to the LHS or already to the HIMS.  

After reaching the highest luminosity, the count rate starts to decrease exponentially, interrupted by several flares. Interestingly, observations associated to flares show slightly higher hardness ratios than those related to the overall decrease. When the source is brightest a total fractional rms of 7.6$\pm$1.1\% is observed, which lies in the 5--10\% range, in which type-B QPOs are expected. According to the studies of MAXI J1659-152 \citep{2011MNRAS.415..292M} and GX 339-4 \citep{2011MNRAS.418.2292M} we would also expect to find type-B QPOs at the time of flares. Although all these points seem to indicate that the source is in the SIMS during these flares, the final prove, namely the detection of type-B QPOs, is missing. Another point which seems to favour that the source entered the SIMS around the brightest observation is the detection of faint radio emission with a possible optically thin spectrum \citep{2011ATel.3364....1M}. This emission is likely corresponding to the variable, quenched radio emission typically seen prior to a radio flare \citep[\eg][]{2004MNRAS.355.1105F}, rather than emission from a steady, compact, flat-spectrum core jet. Regarding the behaviour of the spectral parameters, there is only a correlation between the flares and an increase in the scattered fraction observable. We do not see flares in the inner disc radius, which stays constant, nor in the inner disc temperature. The temperature decrease maps the decrease in accretion rate. After this period of flaring decrease the count rate starts to increase again and enters in another flaring period. During this phase of the outburst the hardness ratios are in the range observed during the previous decay. 

The absence of type-B QPOs makes it impossible to clearly indicate transitions between the SIMS and the HSS. There are several possibilities for this absence. Either the QPO is truly absent, or the amplitude of the QPO is too weak to be detected. In the case of type-B QPOs an amplitude of $\sim$2 -- 4\% rms has to be expected \citep[][]{2005ApJ...629..403C}. For observation \#7 we obtain a 3$\sigma$ upper limit of 4.11\% rms for the amplitude. Therefore, we cannot exclude the possibility that a QPO, which is too weak to be detected, is nevertheless present. A likely reason for the weakness of the QPO might be that the hard emission ($>$ 10 keV) is very faint in this state. We clearly see type-C QPOs when the spectrum is hard-dominated. As soon as the spectrum starts to be disc dominated (namely in the SIMS), we do not detect any QPO. Since QPOs have hard spectra \citep{2006MNRAS.370..405S}, they are related to the hard emission, and thus if the hard emission is faint we will not see the QPOs. The detection of strong type-C QPOs suggests that the lack of type-B QPOs is due to a physical reason, or alternatively, type-B QPOs must be produced in a completely different way with respect to type-C QPOs.
However, we cannot exclude that it might be the lower statistics due to the faintness of the source which permit us to clearly detect type-B QPOs. 
A fact, quite often encountered in studies of extragalactic sources, where count rates are usually rather low compared to Galactic objects, due to the larger distance to the source \citep[see \eg\ Sect.10.3 in][]{2011A&A...534A..55S}. 

Although we cannot clearly indicate when \maxi\ entered the HSS, we know for sure that it reached the HSS on day 58. Our spectral investigation reveals that the source spectra are purely disc dominated after that date. This means that the source spectrum can be described completely by a multi-color disc blackbody model. The emission detected at higher energies can be attributed to the Galactic ridge X-ray emission and is modeled with a power law component. The domination of the spectrum by the disc emission and the constancy of the hard (Galactic ridge X-ray) emission lead to the almost constant hardness ratio during this part of the outburst (c.\,f.\ Fig.\,\ref{Fig:HID}). The hardness ratio remains almost constant since the disc temperature is changing very slowly. More important than the changes in the disc parameters is the flux decrease. To investigate whether the emission detected at higher energies can contain a contribution from the source, which will be difficult to disentangle from the GRXE, due to the faintness of the source, we fitted HSS spectra with a \emph{diskbb + powerlaw} model. This allowed us to determine the flux of the power law component in the 3 -- 20 keV band. The obtained fluxes are a few times $10^{-11}$ erg cm$^2$ s$^{-1}$, which seem to indicate a contribution of the source, as the GRXE is expected to contribute not more than $\sim 10^{-11}$ erg cm$^2$ s$^{-1}$ \citep{2006A&A...452..169R}. We used an RXTE observation of XTE J1752-223 at about the same hardness ratio (95360-01-02-08), in which the power law component is clearly observed and created an artificial spectrum with a count rate reduced by about a factor eight. This artificial spectrum can be well fitted with a \emph{diskbb + powerlaw} model, with a photon index fixed at $\Gamma=$2.1. The flux in the 3 -- 20 keV band associated with the power law component is comparable to that expected from the GRXE. Thus, it seems very likely that even after day 58 \maxi\ emits with a spectrum similar to those of brighter sources at about the same hardness ratio, although at higher energies the emission of the source cannot be disentangled from the GRXE, due to the faintness of the source. 

A striking peculiarity of this outburst of \maxi\ is its faintness. In observation \#7, in which the source is brightest, it reaches a maximum unabsorbed flux of (5.4--5.7)\ergcm{-9} or (0.22--0.24) Crab in the 2 -- 20 keV band, depending on the assumed foreground absorption. Similar flux levels have been observed during the hard to soft transition in the 2004 outburst of GX 339-4, although this source normally shows these transitions at much higher fluxes \citep{2011MNRAS.418.2292M}. In any case, we can exclude that the foreground absorption is responsible for the observed faintness of \maxi. This leaves two other possible explanations. Either the source is located rather far away and hence appears faint, or it is at a similar distance than other black hole binary candidates but is intrinsically faint.
\citet{1998MNRAS.293L..42K} show that the light curves of soft X-ray transients can be described in the disc instability picture used to explain dwarf nova outbursts \citep[see][for a review]{1993adcs.book....6C}, if irradiation by the central X-ray source during the outburst is taken into account. Irradiation prevents that the disc can return to the cool state until the central accretion rate is sharply reduced. If irradiation is strong enough to ionize all of the disc out to its edge, \citet{1998MNRAS.293L..42K} show that the X-ray flux will decay exponentially and that the recurrence time to the next outburst will be long, as the accretion disc has to be rebuild by mass transfer from the companion star. In addition, a thermal instability in the outer disc will be triggered causing a secondary maximum in the light curve about 50 -- 75 days after the start of the outburst. In the case of \maxi\ we observe a secondary maximum in the light curve about 40 days after the start of the RXTE monitoring. This suggests that the outburst started about 10 --- 20 days earlier, which is plausible as RXTE missed to observe the initial flux rise. Ionisation of the whole disc, and hence exponential flux decay, 
is only to be expected if the peak X-ray luminosity is larger than $L_{\mr{crit}}=3.7\times10^{36}R^2_{11}$erg s$^{-1}$, where $R_{11}$ denotes the radius of the ionised disc in units of $10^{11}$ cm \citep{1998MNRAS.301..382S}. As we detected an exponential decay of the count rate with a decay timescale of $\sim$43 days, we can obtain an estimate for the distance of the source. Using the equation $\tau\mr{[days]}\simeq40R^{4/5}_{11}$ \citep{1998MNRAS.293L..42K} together with the measured decay timescale of $\tau\simeq43$ days, we obtain a critical luminosity of  $L_{\mr{crit}}\simeq$4\ergs{36}. The unabsorbed peak flux is $f_{\mr{p}}\simeq$5.7\ergcm{-9} (2--20 keV). This gives a lower limit for the distance of the source of about 8.5 kpc, which is about the distance to GX 339-4 \citep{2004MNRAS.351..791Z}. Similar distances have been also observed for other black hole X-ray binary candidates \citep[see Table 1 in][]{2010MNRAS.403...61D}. Using this estimated distance, the luminosity of the source throughout the outburst is comparable with that of GX 339-4 during outburst decay. This finding suggests that \maxi\ is intrinsically faint. 
Since we only obtained a lower limit for the distance there is still the possibility that the source appears faint because it is located far away. To reach at least during the peak of the outburst a luminosity similar to the one observed during hard-to-soft transitions in GX 339-4 the distance of \maxi\ must be about 12 -- 13 kpc. With such a distance \maxi\ would belong to the group of most distant black hole X-ray binaries within our Galaxy \citep{2010MNRAS.403...61D}.  Thus, we cannot exclude the possibility that \maxi\ is a normally bright, but rather distant source. The available radio data \citep{2011ATel.3364....1M} do not help to obtain the distance, as the spectral index is poorly constrained and hence the detected radio emission might be optically thin.
 
\section{Conclusion}
We presented the results of our X-ray spectral and timing analysis of the black hole candidate \maxi\ during its first observed outburst. The outburst evolution of the system is similar to that known from other black hole candidates, although this outburst was observed at rather low count rate. This hampered a clear determination of state transitions. Special features observed during this outburst of \maxi\ are (i) an exponential decay interrupted by several flares, which took place after the source was brightest, and most probably has to be associated to the SIMS/HSS, (ii) a period of the HSS during which the source spectrum appears purely disc dominated and the count rate decays exponentially. We showed that the source spectrum during this period of the HSS should be similar to spectra of other black hole X-ray binary candidates observed at a similar hardness ratio. It is the low count rate which makes it hard to disentangle the contribution of the hot Comptonized component from the Galactic ridge X-ray emission. The presence of the exponential decay allowed us to estimate that the distance to the source is more than 8.5 kpc. Similar distances have been obtained for other black hole X-ray binary candidates. This implies that the source was observed at low count rate because it is intrinsically faint. However, the possibility that the source emits at a luminosity similar to that of other black hole X-ray binary candidates still remains, if the source has a distance of more than 12 kpc. Complementary results obtained through multiwavelengths campaigns of the present and forthcoming outbursts of this source will result in a deeper understanding of the behaviour observed in this source and of the accretion process taken place in black hole binaries.

\section*{Acknowledgments}
As this will be one of the last BHC outburst observed with RXTE, we would like to use this opportunity to thank the RXTE team for several years of continous monitoring of BH X-ray binaries, which resulted in remarkable insights into the main properties of these objects.
The research leading to these results has received funding from the European Community's Seventh Framework Programme (FP7/2007-2013) under grant agreement number ITN 215212 ``Black Hole Universe". TMD acknowledges support by ERC advanced investigator grant 267697-4PI-SKY. SM and TB acknowledge support from grant ASI-INAF I/009/10/0.
This work makes use of EURO-VO software, tools or services. The EURO-VO has been funded by the European Commission through contracts RI031675 (DCA) and 011892 (VO-TECH) under the 6th Framework Programme and contracts 212104 (AIDA) and 261541 (VO-ICE) under the 7th Framework Programme.

\bibliographystyle{mn2e}
\bibliography{/Users/apple/work/papers/my2010}

\appendix

\section[]{Online Material}
\onecolumn
\begin{center} 
\begin{longtable}{lrrrrlccccc}
\caption{Spectral parameters derived from the best fit for each observation. A model consisting of simpl and diskbb was used. The foreground absorption was fixed at $N_{\mr{H}}=$1.4\hcm{22}.}\label{tab:data}\\
\hline\noalign{\smallskip}
 \multicolumn{1}{c}{\#} &\multicolumn{1}{c}{Obs. id.} &  \multicolumn{1}{c}{day} &  \multicolumn{1}{c}{MJD} & \multicolumn{1}{c}{$\chi^2_{\mr{red}}$} & \multicolumn{1}{c}{diskbb norm} & \multicolumn{1}{c}{Tin} & \multicolumn{1}{c}{$\Gamma$} &  \multicolumn{1}{c}{FracSctr}\\
  & & & & & & \multicolumn{1}{c}{[keV]} & & \\
\hline\noalign{\smallskip} 
\endfirsthead
\caption{continued from previous page}\\
\hline\noalign{\smallskip}
\multicolumn{1}{c}{\#} & \multicolumn{1}{c}{Obs. id.} &  \multicolumn{1}{c}{day} &  \multicolumn{1}{c}{MJD} & \multicolumn{1}{c}{$\chi^2_{\mr{red}}$} & \multicolumn{1}{c}{diskbb norm} & \multicolumn{1}{c}{Tin} & \multicolumn{1}{c}{$\Gamma$} &  \multicolumn{1}{c}{FracSctr}\\
  & & & & & & \multicolumn{1}{c}{[keV]} & &\\
\hline\noalign{\smallskip} 
\endhead
\hline
\noalign{\smallskip}
\endfoot
\hline
\noalign{\smallskip}
\endlastfoot
 1   & 96371-02-01-00    &     0.0 & 55691.1 & 0.86  & $  118.92_{- 68.92}^{ +633.85}    $&$  0.88_{-0.23 }^{+0.17}   $&$  1.78_{-0.07 }^{+0.07}   $&$ 0.752_{-0.051}^{+ 0.070} $\\
 2   & 96371-02-01-01    &     1.0 & 55692.1 &  0.91  & $  285.02_{-205.50}^{+958.76}    $&$  0.75_{-0.16 }^{+0.20}   $&$  1.88_{-0.07 }^{+0.07}   $&$ 0.723_{-0.051 }^{+0.059} $\\
 3   & 96371-02-01-02    &     2.0 & 55693.1 &  0.99  & $  347.48_{-118.20}^{+176.81}    $&$  0.76_{-0.05 }^{+0.06}   $&$  2.01_{-0.04 }^{+0.04}   $&$ 0.620_{-0.021 }^{+0.022} $\\ 
 4   & 96371-02-02-00    &     3.0 & 55694.1 &   1.19  & $  380.05_{-204.17}^{+456.35}    $&$  0.77_{-0.10 }^{+0.11}   $&$  2.21_{-0.10 }^{+0.09}   $&$ 0.549_{-0.049 }^{+0.054} $\\  
 5   & 96371-02-02-01    &     3.8 & 55694.9 &   0.96  & $  383.38_{-116.40}^{+200.46}    $&$  0.79_{-0.05 }^{+0.06}   $&$  2.29_{-0.07 }^{+0.08}   $&$ 0.482_{-0.027 }^{+0.031} $\\ 
 6   & 96371-02-02-02    &     4.6 & 55695.7 &   0.76  & $  288.73_{- 61.31}^{+  120.16}   $&$  0.91_{-0.06 }^{+0.05}   $&$  2.33_{-0.08 }^{+0.08}   $&$ 0.400_{-0.024}^{+ 0.031} $\\ 
 7   & 96371-02-02-03    &     5.6 & 55696.7 &  0.96  & $  361.01_{- 74.02}^{+ 160.49}    $&$  0.89_{-0.06 }^{+0.04}   $&$  2.49_{-0.07 }^{+0.08}   $&$ 0.422_{-0.024}^{+ 0.037} $\\  
 8   & 96371-02-02-04    &     7.0 & 55698.1 &  0.90  & $  281.43_{- 52.06}^{+ 101.98}    $&$  0.93_{-0.05 }^{+0.04}   $&$  2.41_{-0.08 }^{+0.09}   $&$ 0.334_{-0.020}^{+ 0.028} $\\
 9   & 96371-02-02-05    &     8.1 & 55699.2 &  0.94  & $  215.59_{- 26.44}^{+  37.44}     $&$  0.91_{-0.03 }^{+0.03}     $&$  2.30_{-0.17 }^{+0.19}   $&$ 0.158_{-0.013 }^{+0.017} $\\
10  & 96371-02-02-06    &     9.0 & 55700.1 &  0.74  & $  257.44_{- 46.39}^{+  60.65}     $&$  0.93_{-0.04 }^{+0.04}   $&$  2.36_{-0.12 }^{+0.12}   $&$ 0.285_{-0.016}^{+ 0.026} $\\
11  & 96371-02-02-07    &     9.2 & 55700.3 &  0.88  & $  273.82_{- 39.07}^{+  56.78}     $&$  0.89_{-0.03 }^{+0.03}   $&$  2.32_{-0.10 }^{+0.12}   $&$ 0.239_{-0.014}^{+ 0.017} $\\
12  & 96371-02-03-00    &   10.8 & 55701.9 & 1.15   & $  241.79_{- 27.26}^{+  31.70}     $&$  0.90_{-0.02 }^{+0.02}   $&$  2.22_{-0.08 }^{+0.09}   $&$ 0.219_{-0.010 }^{+0.012} $\\
13  & 96371-02-03-01    &   11.7 & 55702.8 & 0.88   & $  240.44_{- 49.50}^{+  91.60}     $&$  0.92_{-0.05 }^{+0.04}   $&$  2.48_{-0.13 }^{+0.16}   $&$ 0.307_{-0.014 }^{+0.044} $\\
14  & 96371-02-03-02    &   12.9 & 55704.0 & 1.44   & $  250.13_{- 26.79}^{+  32.65}     $&$  0.89_{-0.02 }^{+0.02}   $&$  2.22_{-0.08 }^{+0.08}   $&$ 0.218_{-0.007 }^{+0.010} $\\
15  & 96371-02-03-03    &   14.0 & 55705.1 & 0.75   & $  286.65_{- 57.62}^{+ 111.91}    $&$  0.90_{-0.05 }^{+0.04}   $&$  2.37_{-0.11 }^{+0.12}   $&$ 0.333_{-0.023 }^{+0.030} $\\
16  & 96371-02-03-06    &   14.2 & 55705.3 & 1.02   & $  319.42_{- 58.67}^{+  73.92}     $&$  0.89_{-0.03 }^{+0.04}   $&$  2.42_{-0.08 }^{+0.08}   $&$ 0.357_{-0.010 }^{+0.024} $\\
17  & 96371-02-03-04    &   15.2 & 55706.3 & 1.02   & $  327.13_{- 47.14}^{+  88.62}     $&$  0.88_{-0.04 }^{+0.03}   $&$  2.37_{-0.07 }^{+0.08}   $&$ 0.345_{-0.017 }^{+0.020} $\\
18  & 96371-02-03-05    &   16.5 & 55707.5 & 1.19   & $  231.57_{- 25.80}^{+  47.74}     $&$  0.89_{-0.03 }^{+0.02}   $&$  2.16_{-0.09 }^{+0.11}   $&$ 0.210_{-0.009 }^{+0.014} $\\
19  & 96371-02-04-06    &   17.0 & 55708.1 & 1.07   & $  290.38_{- 65.67}^{+  92.16}     $&$  0.85_{-0.04 }^{+0.04}   $&$  2.59_{-0.22 }^{+0.22}   $&$ 0.203_{-0.026 }^{+0.031} $\\  
20  & 96371-02-04-00    &   17.3 & 55708.4 & 0.88   & $  258.77_{- 31.01}^{+  44.57}     $&$  0.87_{-0.03 }^{+0.02}   $&$  2.36_{-0.09 }^{+0.10}   $&$ 0.246_{-0.012 }^{+0.018} $\\
21  & 96371-02-04-01    &   18.4 & 55709.5 & 1.00   & $  251.39_{- 28.07}^{+  34.68}     $&$  0.87_{-0.02 }^{+0.02}   $&$  2.26_{-0.08 }^{+0.08}   $&$ 0.226_{-0.009 }^{+0.012} $\\
22  & 96371-02-04-02    &   19.6 & 55710.7 & 1.05   & $  233.20_{- 26.53}^{+  31.56}     $&$  0.88_{-0.02 }^{+0.02}   $&$  2.13_{-0.09 }^{+0.10}   $&$ 0.202_{-0.009 }^{+0.010} $\\  
23  & 96371-02-04-03    &   20.5 & 55711.6 & 1.42   & $  267.17_{- 34.59}^{+  39.34}     $&$  0.89_{-0.02 }^{+0.02}   $&$  2.27_{-0.06 }^{+0.06}   $&$ 0.322_{-0.011 }^{+0.013} $\\  
24  & 96371-02-04-04    &   21.4 & 55712.5 & 1.00   & $  250.83_{- 29.37}^{+  36.41}     $&$  0.87_{-0.02 }^{+0.02}   $&$  2.28_{-0.09 }^{+0.09}   $&$ 0.245_{-0.012 }^{+0.013} $\\  
25  & 96371-02-04-05    &   23.5 & 55714.6 & 0.96   & $  223.77_{- 21.98}^{+  25.27}     $&$  0.87_{-0.02 }^{+0.02}   $&$  2.08_{-0.11 }^{+0.12}   $&$ 0.154_{-0.006 }^{+0.009} $\\  
26  & 96430-01-01-00    &   24.4 & 55715.5 & 1.06   & $  243.18_{- 25.54}^{+  31.39}     $&$  0.86_{-0.02 }^{+0.02}   $&$  2.18_{-0.10 }^{+0.11}   $&$ 0.183_{-0.008 }^{+0.011} $\\
27  & 96430-01-01-01    &   25.5 & 55716.6 & 0.97   & $  227.60_{- 28.69}^{+  34.46}     $&$  0.86_{-0.02 }^{+0.02}   $&$  2.07_{-0.12 }^{+0.13}   $&$ 0.194_{-0.009 }^{+0.011} $\\
28  & 96430-01-01-02    &   26.7 & 55717.8 & 0.77   & $  243.24_{- 33.76}^{+  42.54}     $&$  0.85_{-0.03 }^{+0.02}   $&$  2.19_{-0.12 }^{+0.15}   $&$ 0.204_{-0.010 }^{+0.018} $\\
29  & 96430-01-01-03    &   27.6 & 55718.7 & 0.65   & $  239.31_{- 33.42}^{+  60.46}     $&$  0.85_{-0.03 }^{+0.02}   $&$  2.24_{-0.14 }^{+0.16}   $&$ 0.192_{-0.011 }^{+0.019} $\\
30  & 96430-01-01-04    &   28.3 & 55719.4 & 1.05   & $  276.35_{- 29.63}^{+  33.49}     $&$  0.84_{-0.02 }^{+0.02}   $&$  2.16_{-0.11 }^{+0.12}   $&$ 0.174_{-0.008 }^{+0.011} $\\  
31  & 96430-01-01-05    &   29.5 & 55720.6 & 0.75   & $  237.05_{- 24.71}^{+  32.67}     $&$  0.85_{-0.02 }^{+0.02}   $&$  2.12_{-0.10 }^{+0.11}   $&$ 0.190_{-0.007 }^{+0.011} $\\
32  & 96430-01-01-06    &   30.3 & 55721.4 & 0.85   & $  220.51_{- 30.84}^{+  43.12}     $&$  0.85_{-0.03 }^{+0.02}   $&$  2.15_{-0.13 }^{+0.15}   $&$ 0.195_{-0.012 }^{+0.015} $\\
33  & 96430-01-02-00    &   32.2 & 55723.3 & 0.96   & $  151.04_{- 17.38}^{+  20.88}     $&$  0.90_{-0.02 }^{+0.02}   $&$  2.02_{-0.09 }^{+0.10}   $&$ 0.223_{-0.008 }^{+0.010} $\\
34  & 96430-01-02-01    &   33.4 & 55724.5 & 0.95   & $  193.10_{- 21.59}^{+  26.01}     $&$  0.87_{-0.02 }^{+0.02}   $&$  2.02_{-0.10 }^{+0.11}   $&$ 0.197_{-0.007 }^{+0.009} $\\
35  & 96430-01-02-02    &   34.2 & 55725.3 & 0.90   & $  262.14_{- 44.88}^{+  58.77}     $&$  0.87_{-0.03 }^{+0.03}   $&$  2.25_{-0.09 }^{+0.09}   $&$ 0.315_{-0.016 }^{+0.019} $\\
36  & 96430-01-02-03    &   35.3 & 55726.4 & 1.39   & $  243.01_{- 33.01}^{+  60.69}     $&$  0.84_{-0.03 }^{+0.02}   $&$  2.10_{-0.12 }^{+0.14}   $&$ 0.218_{-0.010 }^{+0.017} $\\  
37  & 96430-01-02-04    &   36.8 & 55727.9 & 1.00   & $  193.36_{- 28.32}^{+  53.18}     $&$  0.89_{-0.04 }^{+0.03}   $&$  1.76_{-0.22 }^{+0.27}   $&$ 0.136_{-0.006 }^{+0.011} $\\  
38  & 96430-01-03-00    &   38.6 & 55729.7 & 0.64   & $  238.68_{- 25.08}^{+  27.81}     $&$  0.87_{-0.02 }^{+0.02}   $&$  2.00_{-0.17 }^{+0.19}   $&$ 0.111_{-0.006 }^{+0.009} $\\  
39  & 96430-01-03-01    &   39.2 & 55730.3 & 1.35   & $  224.50_{- 20.27}^{+  24.32}     $&$  0.87_{-0.02 }^{+0.02}   $&$  2.01_{-0.12 }^{+0.13}   $&$ 0.129_{-0.005 }^{+0.007} $\\  
40  & 96430-01-03-07    &   39.9 & 55731.0 & 0.69   & $  262.87_{- 35.14}^{+  41.12}     $&$  0.86_{-0.02 }^{+0.02}   $&$  2.10_{-0.22 }^{+0.23}   $&$ 0.123_{-0.009 }^{+0.014} $\\
41  & 96430-01-03-02    &   40.0 & 55731.1 & 0.92   & $  194.22_{- 28.98}^{+  36.00}     $&$  0.89_{-0.03 }^{+0.03}   $&$  2.11_{-0.22 }^{+0.25}   $&$ 0.124_{-0.009 }^{+0.015} $\\  
42  & 96430-01-03-03    &   41.1 & 55732.2 & 1.10   & $  241.11_{- 25.73}^{+  34.17}     $&$  0.88_{-0.02 }^{+0.02}   $&$  1.88_{-0.17 }^{+0.25}   $&$ 0.109_{-0.004 }^{+0.011} $\\
43  & 96430-01-03-04    &   42.3 & 55733.4 & 0.63   & $  235.30_{- 25.92}^{+  31.76}     $&$  0.88_{-0.02 }^{+0.02}   $&$  1.92_{-0.18 }^{+0.20}   $&$ 0.118_{-0.006 }^{+0.009} $\\  
44  & 96430-01-03-05    &   43.3 & 55734.4 & 0.59   & $  287.50_{- 59.91}^{+  87.31}     $&$  0.85_{-0.04 }^{+0.04}   $&$  2.05_{-0.35 }^{+0.36}   $&$ 0.145_{-0.016 }^{+0.029} $\\
45  & 96430-01-03-06    &   44.1 & 55735.2 & 1.67   & $  284.06_{- 27.64}^{+  33.67}     $&$  0.85_{-0.02 }^{+0.02}   $&$  2.32_{-0.08 }^{+0.09}   $&$ 0.196_{-0.009 }^{+0.011} $\\  
46  & 96430-01-04-00    &   45.3 & 55736.4 & 0.65   & $  274.50_{- 30.98}^{+  37.25}     $&$  0.87_{-0.02 }^{+0.02}   $&$  2.35_{-0.07 }^{+0.07}   $&$ 0.257_{-0.011 }^{+0.012} $\\
47  & 96430-01-04-01    &   46.2 & 55737.3 & 0.90   & $  237.56_{- 25.08}^{+  32.94}     $&$  0.87_{-0.02 }^{+0.02}   $&$  2.19_{-0.11 }^{+0.12}   $&$ 0.180_{-0.007 }^{+0.012} $\\  
48  & 96430-01-04-02    &   47.1 & 55738.2 & 0.65   & $  310.92_{- 80.72}^{+  82.07}     $&$  0.84_{-0.03 }^{+0.05}   $&$  2.31_{-0.24 }^{+0.21}   $&$ 0.177_{-0.023 }^{+0.025} $\\
49  & 96430-01-04-03    &   48.1 & 55739.2 & 1.31   & $  290.14_{- 31.76}^{+  43.97}     $&$  0.85_{-0.02 }^{+0.02}   $&$  2.38_{-0.08 }^{+0.08}   $&$ 0.255_{-0.013 }^{+0.015} $\\
50  & 96430-01-04-04    &   49.2 & 55740.3 & 0.98   & $  283.10_{- 47.76}^{+  63.92}     $&$  0.84_{-0.03 }^{+0.03}   $&$  2.36_{-0.14 }^{+0.15}   $&$ 0.204_{-0.017 }^{+0.021} $\\
51  & 96430-01-04-05    &   50.1 & 55741.2 & 0.91   & $  241.64_{- 43.15}^{+  55.57}     $&$  0.87_{-0.03 }^{+0.03}   $&$  2.32_{-0.14 }^{+0.16}   $&$ 0.239_{-0.019 }^{+0.023} $\\
52  & 96430-01-04-06    &   51.8 & 55742.9 & 0.72   & $  300.86_{- 48.97}^{+  64.81}     $&$  0.83_{-0.03 }^{+0.03}   $&$  2.25_{-0.30 }^{+0.37}   $&$ 0.112_{-0.014 }^{+0.026} $\\
53  & 96430-01-05-00    &   52.6 & 55743.7 & 0.65   & $  332.86_{- 57.40}^{+  79.36}     $&$  0.81_{-0.03 }^{+0.03}   $&$  2.48_{-0.33 }^{+0.26}   $&$ 0.138_{-0.024 }^{+0.024} $\\  
54  & 96430-01-05-01    &   53.7 & 55744.8 & 0.51   & $  254.84_{- 33.55}^{+  41.02}     $&$  0.84_{-0.02 }^{+0.02}   $&$  2.23_{-0.30 }^{+0.34}   $&$ 0.076_{-0.008 }^{+0.014} $\\
55  & 96430-01-05-02    &   54.8 & 55745.9 & 0.72   & $  249.45_{- 32.23}^{+  38.71}     $&$  0.83_{-0.02 }^{+0.02}   $&$  2.25_{-0.23 }^{+0.26}   $&$ 0.094_{-0.008 }^{+0.013} $\\
56  & 96430-01-05-03    &   56.6 & 55746.7 & 0.56   & $  227.83_{- 36.58}^{+  49.79}     $&$  0.84_{-0.03 }^{+0.03}   $&$  2.20_{-0.42}^{+0.52}    $&$ 0.085_{-0.012 }^{+0.027} $\\
57  & 96430-01-05-04    &   57.6 & 55748.7 & 0.59   & $  307.74_{- 62.06}^{+ 113.26}    $&$  0.79_{-0.04 }^{+0.03}   $&$  2.35_{-0.74}^{+0.80}    $&$ 0.072_{-0.018 }^{+0.040} $\\
58  & 96430-01-05-05    &   58.8 & 55749.9 & 0.57   & $  239.57_{- 31.77}^{+  39.10}     $&$  0.82_{-0.02 }^{+0.02}   $&$  2.10                                 $&$0.064_{-0.009}^{+ 0.009} $\\
59  & 96430-01-06-00    &   60.1 & 55751.2 & 0.91   & $  254.20_{- 28.28}^{+  33.48}     $&$  0.81_{-0.01 }^{+0.02}   $&$  2.10                                 $&$0.031_{-0.007}^{+ 0.006} $\\
60  & 96430-01-06-01    &   61.0 & 55752.1 & 1.54   & $  216.51_{- 25.09}^{+  30.12}     $&$  0.83_{-0.02 }^{+0.02}   $&$  2.10                                 $&$0.031_{-0.008}^{+ 0.008} $\\
61  & 96430-01-06-02    &   61.8 & 55752.9 & 0.63   & $  276.68_{- 36.40}^{+  43.64}     $&$  0.79_{-0.02 }^{+0.02}   $&$  2.10                                 $&$0.033_{-0.008}^{+ 0.007} $\\
62  & 96430-01-06-03    &   63.1 & 55754.2 & 0.99   & $  217.67_{- 23.78}^{+  34.74}     $&$  0.82_{-0.02 }^{+0.02}   $&$  2.10                                 $&$0.024_{-0.009}^{+ 0.011} $\\
63  & 96430-01-06-05    &   65.1 & 55756.2 & 0.53   & $  230.18_{- 26.17}^{+  31.40}     $&$  0.81_{-0.02 }^{+0.02}   $&$  2.10                                 $&$0.033_{-0.008}^{+ 0.007} $\\
64  & 96430-01-07-00    &   66.3 & 55757.4 & 1.07   & $  231.54_{- 32.76}^{+  41.08}     $&$  0.80_{-0.02 }^{+0.02}   $&$  2.10                                 $&$0.040_{-0.010}^{+ 0.009} $\\
65  & 96430-01-07-01    &   67.2 & 55758.3 & 1.11   & $  220.75_{- 19.77}^{+  23.03}     $&$  0.81_{-0.01 }^{+0.01}   $&$  2.10                                 $&$0.033_{-0.006}^{+ 0.006} $\\
66  & 96430-01-07-02    &   68.1 & 55759.2 & 0.40   & $  246.93_{- 33.59}^{+  38.18}     $&$  0.79_{-0.02 }^{+0.02}   $&$  2.10                                 $&$0.036_{-0.011}^{+ 0.010} $\\
67  & 96430-01-07-03    &   69.4 & 55760.5 & 0.88   & $  266.98_{- 30.53}^{+  46.71}     $&$  0.78_{-0.02 }^{+0.02}   $&$  2.10                                 $&$0.042_{-0.012}^{+ 0.009} $\\
68  & 96430-01-07-04    &   70.3 & 55761.4 & 1.54   & $  261.64_{- 28.25}^{+  45.73}     $&$  0.78_{-0.02 }^{+0.02}   $&$  2.10                                 $&$0.043_{-0.010}^{+ 0.011} $\\
69  & 96430-01-07-05    &   71.2 & 55762.3 & 0.75   & $  229.54_{- 33.70}^{+  41.97}     $&$  0.79_{-0.02 }^{+0.02}   $&$  2.10                                 $&$0.034_{-0.010}^{+ 0.009} $\\
70  & 96430-01-07-06    &   72.4 & 55763.5 & 0.66   & $  225.46_{- 28.68}^{+  34.65}     $&$  0.79_{-0.02 }^{+0.02}   $&$  2.10                                 $&$0.040_{-0.008}^{+ 0.008} $\\
71  & 96430-01-08-00    &   73.2 & 55764.3 & 0.93   & $  227.40_{- 28.26}^{+  34.14}     $&$  0.79_{-0.02 }^{+0.02}   $&$  2.10                                 $&$0.039_{-0.008}^{+ 0.008} $\\
72  & 96430-01-08-01    &   74.3 & 55765.4 & 0.37   & $  263.59_{- 43.10}^{+  49.89}     $&$  0.76_{-0.02 }^{+0.02}   $&$  2.10                                 $&$0.051_{-0.010}^{+ 0.010} $\\
73  & 96430-01-08-02    &   75.0 & 55766.1 & 0.18   & $  278.46_{- 49.35}^{+  66.70}     $&$  0.75_{-0.02 }^{+0.02}   $&$  2.10                                 $&$0.047_{-0.010}^{+ 0.011} $\\
74  & 96430-01-08-04    &   77.3 & 55768.4 & 0.65   & $  243.71_{- 26.60}^{+  30.30}     $&$  0.76_{-0.01 }^{+0.02}   $&$  2.10                                 $&$0.042_{-0.009}^{+ 0.007} $\\
75  & 96430-01-09-00    &   80.1 & 55771.2 & 1.22   & $  225.16_{- 30.25}^{+  37.32}     $&$  0.76_{-0.02 }^{+0.02}   $&$  2.10                                 $&$0.053_{-0.009}^{+ 0.008} $\\
76  & 96430-01-09-01    &   81.2 & 55772.3 & 1.78   & $  257.09_{- 26.23}^{+  31.15}     $&$  0.75_{-0.01 }^{+0.01}   $&$  2.10                                 $&$0.055_{-0.005}^{+ 0.006} $\\
77  & 96430-01-09-02    &   82.6 & 55773.7 & 0.27   & $  247.36_{- 36.32}^{+  45.54}     $&$  0.75_{-0.02 }^{+0.02}   $&$  2.10                                 $&$0.053_{-0.009}^{+ 0.009} $\\
78  & 96430-01-09-03    &   83.0 & 55774.1 & 1.41   & $  223.33_{- 49.04}^{+  63.57}     $&$  0.76_{-0.03 }^{+0.03}   $&$  2.10                                 $&$0.040_{-0.015}^{+ 0.014} $\\
79  & 96430-01-09-04    &   84.2 & 55775.3 & 1.19   & $  278.97_{- 31.11}^{+  37.08}     $&$  0.73_{-0.01 }^{+0.01}   $&$  2.10                                 $&$0.063_{-0.006}^{+ 0.006} $\\
80  & 96430-01-09-05    &   86.0 & 55777.1 & 1.03   & $  251.94_{- 40.60}^{+  54.08}     $&$  0.74_{-0.02 }^{+0.02}   $&$  2.10                                 $&$0.064_{-0.011}^{+ 0.010} $\\
81  & 96430-01-09-06    &   86.8 & 55777.9 & 0.39   & $  267.83_{- 46.86}^{+  63.35}     $&$  0.73_{-0.02 }^{+0.02}   $&$  2.10                                 $&$0.050_{-0.012}^{+ 0.011} $\\
82  & 96430-01-10-00    &   88.0 & 55779.1 & 0.84   & $  217.08_{- 46.53}^{+  60.53}     $&$  0.75_{-0.03 }^{+0.03}   $&$  2.10                                 $&$0.029_{-0.017}^{+ 0.016} $\\
83  & 96430-01-10-01    &   90.7 & 55780.8 & 0.51   & $  229.58_{- 40.04}^{+  48.77}     $&$  0.74_{-0.02 }^{+0.02}   $&$  2.10                                 $&$0.046_{-0.012}^{+ 0.011} $\\
84  & 96430-01-10-02    &   92.9 & 55784.0 & 0.83   & $  238.34_{- 41.72}^{+  51.04}     $&$  0.72_{-0.02 }^{+0.02}   $&$  2.10                                 $&$0.048_{-0.012}^{+ 0.011} $\\
85  & 96430-01-11-00    &   94.6 & 55785.7 & 0.57   & $  298.78_{- 61.34}^{+  77.78}     $&$  0.69_{-0.02 }^{+0.03}   $&$  2.10                                 $&$0.057_{-0.013}^{+ 0.012} $\\
86  & 96430-01-11-01    &   97.9 & 55789.0 & 1.04   & $  284.85_{- 49.81}^{+  68.73}     $&$  0.69_{-0.01 }^{+0.02}   $&$  2.10                                 $&$0.064_{-0.023}^{+ 0.012} $\\
87  & 96430-01-11-03    & 100.9 & 55792.0 & 0.63   & $  268.46_{- 41.83}^{+  45.93 }    $&$  0.69_{-0.02 }^{+0.02}   $&$  2.10                                 $&$0.063_{-0.010 }^{+0.009} $\\
88  & 96430-01-12-01    & 100.9 & 55792.0 & 0.62   & $  274.25_{- 76.14}^{+ 118.80 }   $&$  0.69_{-0.03 }^{+0.04}   $&$  2.10                                 $&$0.067_{-0.020 }^{+0.019} $\\
89 & 96430-01-12-00     & 103.1 & 55794.2 & 0.91   & $  186.11_{- 47.79}^{+  64.84 }    $&$  0.72_{-0.03 }^{+0.03}   $&$  2.10                                 $&$0.067_{-0.018 }^{+0.016} $\\
90  & 96430-01-12-02    & 105.3 & 55796.4 & 0.81   & $  255.33_{- 65.91}^{+  81.91 }    $&$  0.69_{-0.03 }^{+0.03}   $&$  2.10                                 $&$0.064_{-0.031 }^{+0.018} $\\
91  & 96430-01-12-03    & 107.0 & 55798.1 & 0.55   & $  238.39_{- 54.81}^{+  69.82 }    $&$  0.69_{-0.03 }^{+0.02}   $&$  2.10                                 $&$0.060_{-0.027 }^{+0.015} $\\
92  & 96430-01-13-00    & 109.3 & 55800.4 & 0.64   & $  172.40_{- 50.66}^{+  91.42 }    $&$  0.71_{-0.04 }^{+0.04}   $&$  2.10                                 $&$0.042_{-0.042 }^{+0.027} $\\
93  & 96430-01-13-01    & 111.8 & 55802.9 & 0.63   & $  345.69_{-157.33}^{+331.493} $&$  0.64_{-0.06 }^{+0.06}   $&$  2.10                                 $&$0.094_{-0.030}^{+ 0.025} $\\
94  & 96430-01-13-02    & 112.7 & 55803.8 & 0.47   & $  321.40_{-133.66}^{+230.561} $&$  0.64_{-0.04 }^{+0.05}   $&$  2.10                                 $&$0.074_{-0.029}^{+ 0.026} $\\
95  & 96430-01-14-00    & 116.0 & 55807.1 & 0.39   & $  225.33_{- 57.15}^{+  78.53 }    $&$  0.67_{-0.03 }^{+0.03}   $&$  2.10                                 $&$0.072_{-0.016 }^{+0.014} $\\
96  & 96430-01-14-01    & 119.7 & 55810.8 & 0.27   & $  146.02_{- 57.63}^{+   9.07 }     $&$  0.69_{-0.06 }^{+0.06}   $&$  2.10                                 $&$0.060_{-0.057}^{+ 0.042} $\\
97  & 96430-01-15-00    & 127.0 & 55818.1 & 0.52   & $  197.89_{- 73.22}^{+121.40}     $&$  0.63_{-0.04 }^{+0.04}   $&$  2.10                                 $&$0.077_{-0.041 }^{+0.035} $\\
\end{longtable}  
\end{center} 
\twocolumn

\bsp

\label{lastpage}

\end{document}